\documentstyle[12pt,epsf]{article} 

\def\lsim{\mathrel{\lower2.5pt\vbox{\lineskip=0pt\baselineskip=0pt 
           \hbox{$<$}\hbox{$\sim$}}}} 
\def\gsim{\mathrel{\lower2.5pt\vbox{\lineskip=0pt\baselineskip=0pt 
           \hbox{$>$}\hbox{$\sim$}}}}

\def\Ap{A^{\prime}}
\def\Bp{B^{\prime}}

\def\App{A^{\prime\prime}}

\def\k{\kappa}
\def\p{\partial}
\def\R{{\cal R}}
\def\d{\delta}

\def\L{\Lambda}

\def\P{\phi}
\def\Pp{\phi^{\prime}}
\def\Ppp{\phi^{\prime\prime}}

\def\e{\epsilon}
\def\td{\tilde{d}}
\begin{document} 
\begin{flushright}
DPNU-01-24\\ hep-th/0109040
\end{flushright}

\vspace{10mm}

\begin{center}
{\Large \bf 
 Five-Dimensional Warped Geometry with a Bulk Scalar Field}

\vspace{20mm}
 Masato ITO 
 \footnote{E-mail address: mito@eken.phys.nagoya-u.ac.jp}
\end{center}

\begin{center}
{
\it 
{}Department of Physics, Nagoya University, Nagoya, 
JAPAN 464-8602 
}
\end{center}

\vspace{25mm}

\begin{abstract}
 We explore the diversity of warped metric function in five-dimensional
 gravity including a scalar field and a $3$-brane.
 We point out that the form of the function is determined by a parameter
 introduced here.
 For a particular value of the parameter, the warped metric function
 is smooth without having a singularity, and we show that the bulk
 cosmological constant have a upper bound and must be positive and 
 that the lower bound of five-dimensional fundamental scale 
 is controlled by both the brane tension
 and four-dimensional effective Planck scale. 
 The general warp factor obtained here may relate to models inspired by
 SUGRA or M-theory. 
\end{abstract} 

\newpage 
%
%
 \section{Introduction}
 
 Randall and Sundrum proposed the setup of $3$-branes embedded in the 
 five-dimensional theory with warped metric and discussed an alternative 
 explanation of the hierarchy problem in the form of a fine-tuning between 
 the negative bulk cosmological constant and brane tension
 \cite{Randall:1999ee,Randall:1999vf}.
 The warped metric function in the model is an exponential scaling of
 the metric along the fifth dimension compactified on $S^{1}/Z_{2}$
 orbifold.    
 Originally, the introduction of warped metric have been made by Rabakov and
 Shaposhnikov who discussed the cosmological constant problem
 in six-dimensional theory with warped metric containing a singularity
 \cite{Rubakov:1983bz}.
 Recently, the extensions of the Randall-Sundrum scenario are widely
 made \cite{Collins:2001ni,Collins:2001ed,Ito:2001fd}, 
 in particular, several brane world scenarios with bulk scalar field
 are investigated \cite{Collins:2001ni,Kachru:2000xs,Kachru:2000hf,
 Csaki:2000wz,Binetruy:2000wn,Carroll:2001zy}.  
 Moreover, it is expected that this setup may connect with the AdS/CFT 
 correspondence or string theory 
 \cite{Kachru:2000xs,Kachru:2000hf,Nojiri:2000eb,
 Anchordoqui:2001qc,Anchordoqui:2000du,Gherghetta:2001iv}.

 In this paper, we consider an alternative extension of warped metric
 function in the framework of five-dimensional gravity with
 a scalar field $\P$. 
 In ref. \cite{Gherghetta:2001iv}, it was shown that 
 Randall-Sundrum brane-worlds arise form extremal D-brane configuration.
 Moreover, from a field theory point of view these brane-worlds consist
 of a warped geometry with a bulk scalar field, and the dual theories
 turn out to be non-conformal.
 To study the connection between RS model and SUGRA theories,
 it is important to investigate various types of warp factor in the
 setup with a bulk scalar fields.
 The metric taken here corresponds to the most general metric
 appearing to SUGRA theories.
 We show that the form of the warped metric is controlled by a parameter
 introduced here and investigate the behavior of the metric for the 
 general value as well as particular value of the parameter.

 This paper is organized as follows. In section 2, we describe the setup
 and derive the warped metric function and the jump conditions due to
 the existence of a brane. 
 Moreover, we compute the four dimensional effective Planck scale
 by integrating out a fifth dimension.
 In section 3, a conclusion is described. 
%
%
 \section{The Setup}

 The physics of this model is governed by the following action
 \begin{eqnarray}
  S&=&\int d^{5}x\;\sqrt{-G}
                \left\{
                 \frac{1}{2\k^{2}_{5}}\R
                -\frac{1}{2}\left(\nabla\phi\right)^{2}-\L
                \right\}
   +\int d^{4}x\;\sqrt{-g}\;
      \left\{\;-f(\P)\;\right\}\,,
   \label{eq1}
 \end{eqnarray}
 where $\cal R$ and $\L$ is the curvature and the cosmological constant
 in the bulk, respectively.
 Here $1/\kappa^{2}_{5}=M^{3}_{\ast}$ where $M_{\ast}$ is the
 fundamental scale of five-dimensional theory, and 
 $G$ is the five-dimensional metric and $g$ is the induced 
 four-dimensional metric on the brane which is located at $y=0$, 
 where $y$ is the coordinate of fifth dimension.
 In Eq.(\ref{eq1}), the second term denotes the brane tension with 
 dependence of a scalar field.
 We take the following metric
 \begin{eqnarray}
  ds^{2}&=&e^{2A(y)}g_{\mu\nu}dx^{\mu}dx^{\nu}+e^{2B(y)}dy^{2}
  \nonumber\\
  &\equiv&G_{MN}dx^{M}dx^{N}\label{eq2}\,,
 \end{eqnarray}
 where $M,N=0,\cdots,3, 5$ and $g_{\mu\nu}=diag(-,+,+,+)$.
 We shall use the notation $\{x^{\mu}\}$ with $\mu=0,\cdots,3$ for the
 coordinates on the four-dimensional spacetime, and 
 $x^{5}=y$ for the fifth coordinate on an extra dimension.
 Note that the appropriate change of $y$-coordinate can lead to the
 metric of the original Randall-Sundrum model, however, 
 the original metric is not inspired by SUGRA theories. 
 Therefore we can take most general metric Eq.(\ref{eq2})
 appearing in SUGRA theories.

 Using the metric, Einstein equations are given by
 \begin{eqnarray}
  \R_{MN}-\frac{1}{2}G_{MN}\R&=&
  \k^{2}_{5}\left[\;\p_{M}\phi\p_{N}\phi
 -G_{MN}\left\{
  \frac{1}{2}\left(\nabla\phi\right)^{2} +\L
  \right\}\;\right.\nonumber\\
 &&\left.\hspace{1cm}
 -\frac{\sqrt{-g}}{\sqrt{-G}}g_{\mu\nu}\d^{\mu}_{M}\d^{\nu}_{N}f(\P)\d(y)
  \;\right]\,.
  \label{eq3}
 \end{eqnarray}
 While, the equation of motion with respect to a scalar field $\P$
 with $y$-dependence is given by
 \begin{eqnarray}
  \p_{M}\left(\sqrt{-G}G^{MN}\p_{N}\P\right)=
  \sqrt{-g}\;\frac{\p f}{\p\phi}\d(y)
    \label{eq4}\,.
 \end{eqnarray}
 With the metric ansatz, these equations can be written as  
 \begin{eqnarray}
 &&\App + 2\left(\Ap\right)^{2}-\Ap\Bp=
 -\frac{\k^{2}_{5}}{3}\left\{
  \frac{1}{2}(\Pp)^{2}+e^{2B}\L\right\}
 -\frac{\k^{2}_{5}}{3}e^{B}f(\P)\d (y)
  \label{eq5}\,,\\
 &&\left(\Ap\right)^{2}=
 \frac{1}{12}\k^{2}_{5}\left(\Pp\right)^{2}-\frac{\k^{2}_{5}}{6}e^{2B}\L
 \label{eq6}\,,\\
 &&
  4\Ap\Pp-\Bp\Pp+\Ppp
 =e^{B}\frac{\p f}{\p\P}\d(y)\,,\label{eq7}
 \end{eqnarray}
 where the prime represents the derivative with respect to the $y$.

 We take the simple ansatz:
 \begin{eqnarray}
  B=\alpha A \,,\label{eq8}
 \end{eqnarray}
 where $\alpha$ is a parameter at this stage.
 Below, we solve the equation of motion in the bulk and study the warped
 metric function for arbitrary $\alpha$.
 Integrating out the equation in the bulk of Eq.(\ref{eq7}), we have
 \begin{eqnarray}
  \Pp=c\; e^{(\alpha-4)A}\,.\label{eq9}
 \end{eqnarray}
 Hence $c$ is the integration constant and it has mass dimension $[5/2]$.
 Substituting the above equation into Eq.(\ref{eq6}), the equation is 
 expressed as
 \begin{eqnarray}
  \Ap=\e\frac{\sqrt{3}}{6}\k_{5}\left|c\right|
      e^{(\alpha-4)A}\sqrt{1-\frac{2\L}{c^{2}}e^{8A}}\,,
  \label{eq10}
 \end{eqnarray}
 where $\e=\pm$, and the selection of the sign $\e$ determines the 
 branch of the square root.
 Note that this solution make sense when the argument of the square root
 in Eq.(\ref{eq10}) is positive. 

 For $\alpha\neq 4,12$,
 this equation can be obtained by using hypergeometric 
 function\footnote[2]{
 The integral representation of hypergeometric function is given by\\
 $\displaystyle{}_{2}F_{1}(a,b;c;z)=\frac{\Gamma(c)}{\Gamma(b)\Gamma(c-b)}
 \int^{1}_{0}dt\;t^{b-1}(1-t)^{c-b-1}(1-zt)^{-a}\,,
 {\it Re}\; c>{\it Re}\; b>0\,.$\\
 The integral represents a one valued analytic function in the $z$-plane
 cut along the real axis from $1$ to $\infty$.
 } as follows
 \begin{eqnarray}
  e^{(4-\alpha)A}\;
 {}_{2}F_{1}\left(\;\frac{1}{2}\,,\frac{4-\alpha}{8}\,;
 \frac{12-\alpha}{8}\,;\,\frac{2\L}{c^{2}}e^{8A}\;\right)
 =\e\frac{\sqrt{3}}{6}\k_{5}\left|c\right|(4-\alpha)
 \left(y+d\right)\,,
 \label{eq11}
 \end{eqnarray}
 where $d$ is the integration constant and $e^{A(0)}$ can be
 normalized to be unity.
 Note that the above equation is solution to be consistent with
 Eq.(\ref{eq5}), consequently.
 In the case of $\alpha=4,12$, we cannot use the integral representation of 
 hypergeometric function due to vanishing of second or third argument.
 Later, we describe the case of $\alpha=4$.
 For $4<\alpha<12$, since the second argument in the hypergeometric
 function is negative, it becomes positive by using the transformation
 formulas of the function.  
 In the case of $\alpha>12$, both the second and the third argument
 become negative.
 The simplest warped metric of $\alpha=0$ have been already investigated
 in Ref. \cite{Collins:2001ni,Collins:2001ed}.
 Note that the value of $\alpha$ determines the form of the metric 
 function $A(y)$.
 Furthermore, the solution of Eq.(\ref{eq11}) is well-defined on one
 side of $y=-d$ due to ${}_{2}F_{1}\geq 0$
 ( depending on the sign of $\e(4-\alpha)$ )
 \cite{Kachru:2000xs,Kachru:2000hf}. 
 
 The existence of a brane at $y=0$ leads to the jump conditions 
 with respect to the first derivative of $A$ and $\P$.
 From Eqs.(\ref{eq5}) and (\ref{eq7}), the jump conditions are
 \begin{eqnarray}
  \Ap(0+)-\Ap(0-)&=&-\frac{1}{3}\k^{2}_{5}e^{\alpha A(0)}f(\P(0))\,,
  \nonumber\\
  \Pp(0+)-\Pp(0-)&=&e^{\alpha A(0)}\frac{\p f}{\p \P}(\P(0))\,.
  \label{eq12}
 \end{eqnarray}
 We denote the sign and the integration constants for the positive
 region $(y>0)$ by $\e_{+}\,,c_{+}\,,d_{+}$ and those for the negative 
 region $(y<0)$ by $\e_{-}\,,c_{-}\,,d_{-}$.
 Imposing the fact with $A(0)=0$, the jump conditions yield
 \begin{eqnarray}
  \e_{+}\frac{1}{\td_{+}}-\e_{-}\frac{1}{\td_{-}}
  &=&\frac{\p f}{\p\P}(\P(0))\,,\nonumber\\
  \e_{+}\sqrt{\frac{1}{\td^{2}_{+}}-2\L}-
  \e_{-}\sqrt{\frac{1}{\td^{2}_{-}}-2\L}&=&
  -\frac{2}{\sqrt{3}}\k_{5}f(\P(0))\,,\label{eq13}
 \end{eqnarray}
 where 
 \begin{eqnarray}
  \frac{1}{\td_{\pm}}=\frac{6}{\sqrt{3}\k_{5}(4-\alpha)}\;
  {}_{2}F_{1}\left(\;\frac{1}{2}\,,\frac{4-\alpha}{8}\,;
 \frac{12-\alpha}{8}\,;\,\frac{2\L}{c^{2}_{\pm}}\;\right)
 \frac{1}{d_{\pm}}\,.\label{eq14}
 \end{eqnarray} 
 Furthermore, if the fifth dimension is assumed to have certain
 symmetry, the sign $\e_{\pm}$ and the integration constants 
 $c_{\pm},d_{\pm}$ are related each other.
 
 If this setup arises from string theory,
 it is natural that
 a scalar field $\P$ in the bulk is regarded as dilaton field. 
 Namely, in the framework of perturbative string theory,
 $f(\P)$ corresponds to the tree-level dilaton coupling
 with dependence of exponential type $( f(\P)\propto e^{a\P})$.
 Thus, the jump conditions of Eq.(\ref{eq13}) gives the information
 on the general form of dilation coupling \cite{Kachru:2000xs}.
 As for this point, we discuss elsewhere.

 For $\alpha=4$, we can directly solve the differential equation in
 Eq.(\ref{eq10}), the warped function with $Z_{2}$ symmetry is given by
 \begin{eqnarray}
  e^{8A_{\pm}(y)}=
  \frac{c^{2}}{2\L}\left\{
  1-\tanh^{2}\frac{2}{\sqrt{3}}\k_{5}|c|(y\pm d)
  \right\}\,,\label{eq15}
 \end{eqnarray}
 where plus ( minus ) corresponds to the function for $y>0$
 ( $y<0$ ) region.
 Note that this warped metric function is smooth without having
 a singularity.
 The normalization condition $A(0)=0$ gives the constraint on the
 integration constants
 \begin{eqnarray}
  \tanh^{2}\frac{2}{\sqrt{3}}\k_{5}|c|d=1-\frac{2\L}{c^{2}}\,,\label{eq16}
 \end{eqnarray}
 therefore, the allowed range of the bulk cosmological constant is
 \begin{eqnarray}
  0<\L\leq \frac{c^{2}}{2}\,,\label{eq17}
 \end{eqnarray}
 where we assume $\L\neq 0$.
 From Eq.(\ref{eq12}), the jump condition of $\Ap$ yields the relation between
 the bulk cosmological constant and the brane tension
 \begin{eqnarray}
  V=\frac{\sqrt{3(c^{2}-2\L})}{\k_{5}}\,,\label{eq18}
 \end{eqnarray}
 where $f(\P(0))=V$.

 It assumes that the fifth dimension is noncompact.
 Integrating out $y$ in the action Eq.(\ref{eq1}), 
 the resulting four dimensional effective Planck scale $M_{\rm Pl}$ is
 finite
 \begin{eqnarray}
  M^{2}_{\rm Pl}
  =\frac{1}{\k^{2}_{5}}\int^{\infty}_{-\infty}\;dy\;
  \left. e^{(2+\alpha) A}\right|_{\alpha=4}=
  \frac{\sqrt{3}}{2\k^{3}_{5}|c|}\;
  {}_{2}F_{1}\left(\;\frac{3}{4}\,,\frac{1}{2}\,;\frac{7}{4}\,;
  \frac{2\L}{c^{2}}\;\right)\,.\label{eq19}
 \end{eqnarray} 
 Taking account for the range of $\L$ in Eq.(\ref{eq17}), since 
 $1<{}_{2}F_{1}(\frac{3}{4},\frac{1}{2};\frac{7}{4};z)\leq
 \frac{\Gamma(\frac{1}{2})\Gamma(\frac{7}{4})}{\Gamma(\frac{5}{4})}
 \sim 1.79721$ for $0<z\leq 1$,
 we obtain
 \begin{eqnarray}
  M^{2}_{\rm Pl}\lsim \frac{M^{9/2}_{\ast}}{|c|}\,,\label{eq20}
 \end{eqnarray}
 where we used $1/\k^{2}_{5}=M^{3}_{\ast}$.
 From Eqs.(\ref{eq18}) and (\ref{eq20}), the integration constant
 can be eliminated, and the bulk cosmological constant is expressed as
 \begin{eqnarray}
  2\L\lsim \frac{M^{9}_{\ast}}{M^{4}_{\rm Pl}}-\frac{V^{2}}{3M^{3}_{\ast}}
  \label{eq21}\,,
 \end{eqnarray}
 imposing the fact that $\L$ is positive, we have
 \begin{eqnarray}
  M_{\ast}>V^{\frac{1}{6}}M^{\frac{1}{3}}_{\rm Pl}
 \label{eq22}\,.
 \end{eqnarray}
 Thus, the lower bound of the five dimensional fundamental scale
 is controlled by both the brane tension and four dimensional effective
 Planck scale.
 For instance, if the value of brane tension $V$ is TeV scale, 
 we then obtain
 \begin{eqnarray}
  M_{\ast}> 10^{8}\;{\rm GeV}\,.
 \end{eqnarray}
 Consequently, the lower bound of $M_{\ast}$ decreases as the value of 
 the brane tension decreases.
 Especially, in the case of vanishing brane tension ( $V=0$ ), 
 if we take $M_{\ast}\sim 1\;{\rm TeV}$, the inequality Eq.(\ref{eq21}) 
 becomes
 \begin{eqnarray}
  \L\lsim \left(\;1\;{\rm eV}\;\right)^{5}\,,\label{eq23}
 \end{eqnarray}
 therefore, Eq.(\ref{eq18}) leads that $c^2\lsim 2({\rm eV})^{5}$.

 As for the value of $\alpha=12$, solving Eq.(\ref{eq10}),
 since the exact form of the warped metric function is complicated,
 we are going to investigate it in next paper \cite{Ito:2001gk}.
 Furthermore, whether the effective four-dimensional Planck scale is 
 finite depends on the value of $\alpha$ \cite{Ito:2001gk}.
%
%
 \section{Conclusion}

 Under the conjecture that Randall-Sundrum model arises from
 the underlying theories (string or M-theory),
 we showed that warp factors has various types in the setup with a bulk
 scalar field. 
 Namely, we explore the diversity of the five-dimensional warped
 geometry, where
 the warped metric function with $y$-dependence of four-dimensional 
 spacetime metric part and the one of fifth dimension metric part 
 are tied through a parameter $\alpha$.
 The metric taken here corresponds to the metric appearing in SUGRA
 theories to be low energy effective theories of D-brane configurations.
 For arbitrary $\alpha$, the form of the warped metric function is implicitly
 described in terms of $y$ by using a hypergeometric function.
 Taking a specific value of $\alpha=4$, the warped metric
 function without having a singularity can be explicitly obtained, and
 we pointed out that the resulting four-dimensional effective Planck
 scale is finite and that the lower bound of five dimensional
 fundamental scale is determined by the brane tension.
 Furthermore, there exist the allowed bound on the bulk cosmological
 constant.
 Although we concretely investigate a simple case of $\alpha=4$ as mentioned
 above, it is necessary to explore the behavior of the warped metric
 function for various $\alpha$ \cite{Ito:2001gk}. 
 However, it is unknown whether the value of $\alpha$ corresponds to
 concrete types of D-brane configuration.
 As for this point, we are going to describe elsewhere in future.
 As preliminary study of connection between RS model and SUGRA theories,
 the general results obtained here are important.
%
 
\end{document}